\documentclass[10pt,reprint,aip,jcp,twocolumn,floats,english,superscriptaddress]{revtex4-1}
\usepackage{amsthm}
\usepackage{epsf}
\usepackage{dsfont}
\usepackage[T1]{fontenc}
\usepackage[latin9]{inputenc}
\usepackage{lmodern}
\usepackage{geometry}
\usepackage{caption,subcaption}
\usepackage{amsmath}
\usepackage{amssymb}
\usepackage{graphicx}
\usepackage{color}
\usepackage{chngpage}

\usepackage{multirow}

\makeatletter
\@ifundefined{textcolor}{}
{%
 \definecolor{BLACK}{gray}{0}
 \definecolor{WHITE}{gray}{1}
 \definecolor{RED}{rgb}{1,0,0}
 \definecolor{GREEN}{rgb}{0,1,0}
 \definecolor{BLUE}{rgb}{0,0,1}
 \definecolor{CYAN}{cmyk}{1,0,0,0}
 \definecolor{MAGENTA}{cmyk}{0,1,0,0}
 \definecolor{YELLOW}{cmyk}{0,0,1,0}
 }

\makeatother

\usepackage{babel}

\begin{document}
\begin{flushright}
DAMTP-2014-40
\end{flushright}

\title{An Inversion-Relaxation Approach for Sampling Stationary Points of Spin Model Hamiltonians}

\author{Ciaran Hughes}
\email{ch558@cam.ac.uk}
\affiliation{The Department of Applied Mathematics and Theoretical Physics, The
University of Cambridge, Clarkson Road, Cambridge CB3 0EH, UK.}

\author{Dhagash Mehta}
\email{dbmehta@ncsu.edu}
\affiliation{Department of Mathematics, North Carolina State University, Raleigh, NC 27695, USA.}
\affiliation{Department of Chemistry, The University of Cambridge, Lensfield Road, Cambridge CB2 1EW, UK}

\author{David J.~Wales}
\email{dw34@cam.ac.uk}
\affiliation{Department of Chemistry, The University of Cambridge, Lensfield Road, Cambridge CB2 1EW, UK.}

\begin{abstract}
Sampling the stationary points of a complicated potential energy landscape is
a challenging problem.
Here we introduce a sampling method based on relaxation from stationary points of the 
highest index of the Hessian matrix. We illustrate how this approach can
find all the stationary points for potentials or Hamiltonians bounded from above, which includes a large class of important 
spin models, and we show that it is far more efficient than previous methods.
For potentials unbounded from above, the relaxation part of the method is still
efficient in finding minima and transition states, which are usually the primary
focus of attention for atomistic systems.
\end{abstract} 

\maketitle

\section{Introduction} 
Finding stationary points (SPs) of a multivariate non-linear function is a 
frequently arising problem in many areas of science. For example,
locating SPs provides the foundations for global optimisation,\cite{lis87,walesd97a,waless99} thermodynamic
sampling to overcome broken ergodicity,\cite{BogdanWC06,SharapovMM07,SharapovM07,Wales2013} and 
rare event dynamics\cite{Wales02,Wales04,Wales06,BoulougourisT07,XuH08,TsalikisLBT10,LempesisTBTCS11,TerrellWCH12} within the
framework of potential energy landscape theory.\cite{Wales03}
Knowledge of the SPs of the potential energy function,
$V({\bf{x}})$, with ${\bf x}=(x_{1}, \dots, x_{N})$, which is usually a real-valued
function from $\mathbb{R}^{N}$ to $\mathbb{R}$, can be exploited to analyse
the properties of a diverse range of physical systems.\cite{Wales03,RevModPhys.80.167}

SPs are defined as the simultaneous
solutions of the system of equations $\partial V({\bf{x}})/\partial x_{i} = 0$, for all $i=1,\dots, N$.
They can be further classified using the second derivative, or Hessian,
matrix. A SP is a minimum if all the non-zero eigenvalues of the Hessian matrix
evaluated at the SP are positive. The minimum at which
$V({\bf{x}})$ attains its lowest possible value is the global minimum, and the
others are local minima. 
A SP is defined as a saddle of index $i$ if
exactly $i$ non-zero eigenvalues of the Hessian evaluated at the SP are
negative. A SP
corresponds to a non-Morse or singular SP\cite{Gilmore81} if the Hessian matrix evaluated at the SP
has at least one additional zero eigenvalue, after removing the zero
eigenvalues corresponding to global symmetries of $V({\bf{x}})$, such as translation and
rotation.

The stationary equations are usually nonlinear for chemical and
physical systems arising in nature, so it is difficult to find all the SPs
in such cases. There are only a few systems for which all the SPs are known
analytically, e.g.~the one-dimensional nearest-neighbour XY model with
periodic\cite{Mehta:2010pe,vonSmekal:2007ns} and
anti-periodic boundary conditions.\cite{Mehta:2009,vonSmekal:2007ns} In some
cases, only a class of solutions is known analytically, e.g.~for the
two-dimensional nearest-neighbour XY model, a class of solutions 
built upon the corresponding one-dimensional model are known exactly, though
many other solutions may also exist.\cite{PhysRevLett.106.057208,Nerattini:2012pi} Similarly, for the mean-field
XY model, all the solutions for a specific choice of the external magnetic
field term are known.\cite{Casetti:June2003:0022-4715:1091}

When analytical results do not exist, one has to employ a numerical
method to find SPs. However, numerical methods to solve nonlinear equations
come with problems of their own. One of the traditional methods to
solve nonlinear equations is the Newton-Raphson approach, in which one tries to
refine initial guesses to find numerical solutions, up to a chosen
numerical precision, of the given system. The method has a few major drawbacks:
first, no matter how many 
solutions are obtained, there is no guarantee that all of them will be found.
In addition, the solutions with large basins of attraction\cite{mezey81b,mezey87,Wales03} may be repeatedly
found for different random initial guesses, wasting computational resources.

Alternatively, the gradient-square minimisation method
has sometimes been employed,\cite{angelani2000saddles,broderix2000energy} in which one minimises the sum of
the squares of the stationary equations of the given potential, i.e.~$W =
\sum_{i=1}^{N}\big(\partial V({\bf x})/\partial {\bf x}_{i}\big)^2$, using
traditional numerical minimisation methods such as conjugate gradient.
The minima of $W$ with $W=0$ are the desired SPs
of  $V({\bf{x}})$. However, it has been shown\cite{DoyeW02,doye2003comment} 
that the number of minima with
$W>0$, which are not the SPs of $V({\bf{x}})$, outweighs the number of minima
with $W=0$, and so this approach is very inefficient. 
Furthermore, these non-stationary points have an additional zero Hessian eigenvalue,
making the minimisation ill-conditioned.\cite{DoyeW02,WalesD03}
However, a systematic approach based on eigenvector-following can locate
saddles of arbitrary index quite efficiently.\cite{DoyeW02,WalesD03}
This approach is implemented in our {\tt OPTIM} program, which includes
a wide variety of geometry optimisation techniques for locating stationary points
and analysing pathways.
The most efficient minimiser in {\tt OPTIM}\cite{AsenjoSWF13} is probably a modified version of the limited-memory
Broyden--Fletcher--Goldfarb--Shanno (LBFGS) algorithm.\cite{Nocedal80,lbfgs}
Single- and double-ended\cite{TrygubenkoW04} transition state searches are also
implemented, with a selection of gradient-only and second derivative-based eigenvector-following\cite{Wales92,Wales93d}
and hybrid eigenvector-following methods.\cite{munrow99,kumedamw01}

Recently, an approach based on algebraic geometry, called the numerical polynomial
homotopy continuation (NPHC) method, has been introduced to explore the
potential energy landscapes of various models with polynomial-like nonlinearity.\cite{Mehta:2009,Mehta:2011xs}
The basic strategy of the method is as follows:
first, an upper bound on the number of solutions of the given stationary
equations is estimated, usually based on the degrees of each equation; then a
different system consisting of the same variables, having exactly the same
number of solutions as the upper bound on the number of solutions, and is easy
to solve, is constructed. Finally, the new system and the original system are
homotopically connected and each solution of the new system is tracked towards
the original system. While tracking the solutions, some paths may diverge to
infinity and the solutions that reach the original system are then the desired
SPs of the given potential. In this way, it is guaranteed to find  \textit{all}
the SPs. The reader is referred to Refs.~\cite{Mehta:2011xs,Mehta:2011wj} for
details of the precise procedures for constructing the homotopy between the two
systems and path-tracking and Refs.~\cite{Mehta:2009,Mehta:2009zv,Mehta:2011xs,Mehta:2011wj,Kastner:2011zz,Maniatis:2012ex,Mehta:2012wk,Hughes:2012hg,Mehta:2012qr,MartinezPedrera:2012rs,
He:2013yk,Mehta:2013fza,Greene:2013ida,SW:05} for its applications to various areas of physics and chemistry. 

Another approach is the interval method, which also guarantees to find all the SPs
of a given $V(\textbf{x})$. This approach can deal with any nonlinear potential,
including those with non-polynomial nonlinearities.\cite{gwaltney2008interval} 
However, it has only proved successful for a
very small number of atoms and SPs so far.
It is based on bisection searches over the accessible coordinate space, so
it scales poorly with system size.

We note that all these numerical methods share a common problem. Once a
numerical solution is found, one may still want to rigorously verify if it
corresponds to the actual solution. In Ref.~\cite{Mehta:2013zia}, a method is
presented to \textit{certify} whether a numerical solution is indeed in the
quadratic convergence region of an actual SP of the potential, or whether it is in the
linear, or even worse, in the chaotic convergence region.

In the present contribution, we present a new method, the inversion-relaxation approach, which 
is applicable to potentials that are bounded from above, such as spin models.
This framework can be far more efficient than previous methods, although
it is possible that some stationary points could still evade discovery unless
the procedure is run with some extra degree of stochasticity
in the choice of geometry optimisation parameters, such as step sizes.

\begin{table*}[htbp]
\caption{Total CPU time in the format hours:minutes:seconds for the XY model with 
periodic boundary conditions.
PBS stands for the Portable Batch System job scheduler.
}
\begin{tabular}{|c||c|c|c|}
\hline
\multicolumn{1}{|c||}{~~~~~$N$~~~~~~} & ~~ Optimisation Method ~~ &
~~ Number of SP's found ~~& ~~ Total CPU time from PBS ~~ \\\hline \hline
\multirow{3}{*}{$10$ } & RFI & $18$ & $00:00:48$  \\
\cline{2-4}   & Relaxation & $18$ & $00:00:59$ \\
\cline{2-4} & Random Search & $18$ & $00:05:12$  \\ \hline \hline
\multirow{3}{*}{$15$ } & RFI & $33$ & $00:02:47$ \\
\cline{2-4} & Relaxation & $33$ & $00:02:51$    \\
\cline{2-4} & Random Search & $32$ & $03:47:56$  \\ \hline \hline
\multirow{3}{*}{$25$ }   & RFI & $78$ & $00:49:48$ \\
\cline{2-4} & Relaxation & $78$ & $01:15:16$ \\
\cline{2-4} & Random Search & $78$ & $12:52:32$  \\ \hline \hline
\multirow{3}{*}{$30$ } & RFI  & $110$ & $02:24:32$ \\
\cline{2-4}  & Relaxation & $110$ & $02:18:53$ \\
\cline{2-4} & Random Search & $102$ & $21:42:14$  \\ \hline \hline
\multirow{3}{*}{$50$ } & RFI  & $228$ & $34:32:05$ \\
\cline{2-4}  & Relaxation & $227$ & $35:22:52$ \\
\cline{2-4} & Random Search & $218$ & $46:45:17$  \\ \hline \hline
\multirow{3}{*}{$3 \times 3$ } & RFI  & $21$ & $00:02:03$ \\
\cline{2-4}  & Relaxation & $21$ & $00:02:06$ \\
\cline{2-4} & Random Search & $21$ & $00:14:15$  \\ \hline
\end{tabular}
\label{tab:timing}
\end{table*}

\begin{table*}[htbp]
\caption{Total CPU time required to perform relaxation from index
  three SPs in order to find transition states and minima. 
For the random search timings presented here, only transition states
and minima were found. Format in hours:minutes:seconds for the XY model with periodic boundary conditions.
PBS stands for the Portable Batch System job scheduler.
}
\begin{tabular}{|c||c|c|c|}
\hline
\multicolumn{1}{|c||}{~~~~~$N$~~~~~~} & ~~ Optimisation Method ~~ &
~~ Total number of SP's, transition states and minima found ~~& ~~ Total CPU time from PBS ~~ \\\hline \hline
\multirow{3}{*}{$5 \times 5$ } & Relaxation & $80,3,3$ & $00:06:13$  \\
\cline{2-4} & Random Search & $5,3,2$ & $01:11:59$  \\ \hline \hline
\multirow{3}{*}{$6 \times 6$ } & Relaxation & $197,4,4$ & $00:31:29$ \\
\cline{2-4} & Random Search & $7,4,3$ & $01:30:09$  \\ \hline \hline
\multirow{3}{*}{$9 \times 9$ } & Relaxation & $319,17,8$ & $04:35:29$ \\
\cline{2-4} & Random Search & $18,12,6$ & $09:53:12$  \\ \hline 
\end{tabular}
\label{tab:timing_ts}
\end{table*}

\section{Random Search Approach} 
We first describe the random search approach. 
Here, one optimises a
random starting configuration in order to find an SP of index $i$. 
The optimisation is performed using a hybrid
Newton-Raphson/eigenvector-following algorithm, which is available in the {\tt OPTIM} program.
This framework has been employed in previous work to analyse the
energy landscapes of some model structural glass-forming systems.\cite{WalesD03}
The step along eigendirection $\alpha$ is taken
as\cite{walesw96,walesdmmw00}
\begin{equation}
x_\alpha = {\displaystyle -2 g_\alpha \over
\displaystyle \varepsilon^2_\alpha(1+\sqrt{1+4 g_\alpha^2 /\varepsilon^4_\alpha})},
\end{equation}
where $g_\alpha$ and $\varepsilon^2_\alpha$ are the component of the gradient and Hessian eigenvalue
corresponding to eigenvector $\alpha$, respectively.\cite{WalesD03}
The sign of $\varepsilon^2$ that determines whether the step in direction $\alpha$ raises or lowers the energy,
and the standard Newton-Raphson procedure can lead to a stationary point of any index.\cite{Wales92,Wales93d,Wales03}
A small number of Newton-Raphson steps are used in combination with a 
trust radius scheme for the maximum step size,\cite{Fletcher80,simonsjto83,walesw96}
before switching to eigenvector-following with a fixed number of uphill directions
corresponding to the required Hessian index of the saddle.\cite{WalesD03}
Here the steps are taken as
\begin{equation}
x_\alpha = {\displaystyle \pm2 g_\alpha \over
\displaystyle |\varepsilon^2_\alpha|(1+\sqrt{1+4 g_\alpha^2 /\varepsilon^4_\alpha})},
\end{equation}
plus for uphill and minus for downhill.

We denote the parameter $n_{\rm RS}$ as the number of 
random starting configurations for each index of the potential. 
If the maximum Hessian index is $i_{\rm max}$, then we perform 
$n_{\rm RS}\,i_{\rm max}$ optimisations in total.

We judge successful convergence in {\tt OPTIM} using a tolerance of $10^{-10}$ 
for the root-mean-square (RMS) 
force and requiring that the maximum unscaled step falls 
below $10^{-7}$. After a successful optimisation, the energy
of the obtained SP is compared against the list of currently known values. If this
energy has been found before, we move on to the next iteration of the
enumeration loop. A tolerance on energy differences of $10^{-5}$ was
found to be optimal. If a SP with the latest energy value
has not been found before, then the SP
configuration, energy and eigenvalues are saved for further analysis.
In principle, we can repeat this process for different starting seeds and
different values of $n_{\rm RS}$ to enumerate a list of all
SPs for a particular potential.

For the random search method, there is no guarantee that we will find all SPs 
and many starting configurations end up in the same basin of attraction after optimisation. 
Additionally, the efficiency of this method is highly dependent on the generation of pseudo-random starting configurations. 
The naivety of the random search approach reflects the fact that it does not take advantage 
of any properties of the PEL or the connection between SPs of different indices. 
We use these properties to describe a new method that can find all the SPs more efficiently 
for suitable potentials.

\section{Relaxation}
One phase of the enumeration approach corresponds to {\it relaxation\/}.
Here, we start by locating a saddle with the highest possible Hessian index
(a maximum)  using the
hybrid Newton-Raphson/eigenvector-following scheme implemented in {\tt OPTIM}. 
The relaxation approach exploits the fact that
every SP of index $i$ is embedded in the configuration space of a saddle of index $i+1$.
Starting at
a  saddle, we can apply a small perturbation and follow a steepest-descent path downhill
in energy to a SP of the next lowest index. 
We repeat the relaxation for all energy distinct
lower index saddles until we reach a minimum. 
This procedure exploits a generalisation of the Murrell-Laidler theorem for the
energetic ordering of saddles.\cite{murrelll68}

We denote $n_{\rm RM}$ as the number of random starting configurations 
for the relaxation method, but with this approach the random configurations are only used
to find maxima of the potential. 
We perform $n_{\rm RM}$ optimisations to obtain maxima, 
then relax each of the energy distinct maxima obtained.

The systematic relaxation procedure depends on the eigenvector to be followed downhill and the magnitude of the perturbation away from the SP. 
For a SP of index $i$, there are $i$ eigenvectors corresponding to a negative Hessian eigenvalue to be followed downhill,
and two distinct directions in each case. 
The displacements can be arranged as
$\pm i,\pm (i-1), \ldots ,\pm 2, \pm 1$, with $1$  meaning the softest mode
with the smallest magnitude negative eigenvalue, $2$
meaning the next softest, etc. The plus/minus signs allow for perturbations in
opposite directions along the same eigenvector. 
The eigenvector to be followed downhill in energy, $v_{\rm down}$, is  
specified, along with the magnitude of the initial step, $\delta$,
which is taken to perturb the system along this Hessian eigendirection.
In principle, this procedure should find all SPs. 

\subsection{Relaxation Following Inversion} 
The relaxation following inversion (RFI) approach differs from the relaxation method only in how we obtain
the initial maximum that we wish to relax. 
In the RFI procedure we employ minimisation for the inverted landscape $-V({\bf x})$.
Since minimisation is generally significantly faster than saddle searches,
the inverted potential provides a much more efficient route to the saddles with
the highest possible index, corresponding to the number of degrees of freedom
that remain after allowing for global symmetries that produce zero eigenvalues.
In the current formulation it requires the potential to be bounded
from above, since otherwise the effective maximisation will generally diverge,
for example, by atom clashing. However, it may be possible to work around this issue by reformulating the (physically irrelevant) regions to avoid singularities.
On the other hand, almost all the continuous spin model Hamiltonians (e.g., the XY model, the N-vector models, etc.) are bounded from above,
providing a wide range of important applications, including glassy landscapes.

We observe that the RFI approach should be efficient for its intended purpose,
namely finding all the SPs. The computational resources required are
likely to increase rapidly with the number of degrees of freedom involved.
If we only want to sample 
a subset of SPs, then the random search method may be more convenient.
Nevertheless, one could start relaxation from a saddle 
of index $j>1$ to sample transition states (index one saddles\cite{murrelll68}) and minima. 

The following steps provide an overview of our implementation:
\begin{enumerate}
\item Choose values for the parameter $\delta$ and the maximum
permitted step size, $\Delta$. These parameters are then fixed
throughout the entire relaxation process.
\item Find a maximum of the potential using minimisation for the inverted potential.
\item Relax the maximum along all downhill eigenvectors 
using hybrid Newton-Raphson/eigenvector-following in order to find SPs of the next lowest index.
\item Recursively relax all energy distinct saddles found, ignoring any duplicates, along 
all downhill eigenvectors. 
\item Repeat steps 2-4 as necessary.
\end{enumerate}

\section{Numerical Experiments} 
We compare the methods discussed in the previous section for a specific model, 
the XY model without disorder, which has attracted considerable interest 
in statistical mechanics
in recent years. 
The model is important because it is one of the simplest
lattice spin systems with continuous configuration space, unlike the Ising model. 
It is employed in studies of superfluid helium, low temperature superconductivity, 
Josephson junction arrays and hexatic liquid crystals due 
to its rich energy landscape and dynamics, including its unique
phase transition properties \cite{kosterlitz1973ordering} in 2D. 
The same model is also used in particle physics as 
the lattice Landau gauge functional for a compact
$U(1)$ lattice gauge theory.\cite{Maas:2011se,Mehta:2009,Mehta:2010pe,vonSmekal:2007ns,vonSmekal:2008es,Mehta:2009zv,Hughes:2012hg,Mehta:2014jla}
Moreover, it appears in the complex systems field as
the nearest-neighbor Kuramoto model with homogeneous
frequency. There, the SPs of the model constitute special configurations
in phase space from a non-linear dynamical systems viewpoint.\cite{acebron2005kuramoto}

The XY model Hamiltonian can be written as:
\begin{equation}
V=\frac{1}{N^d}\sum_{j=1}^{d}\sum_{\textbf{i}}[1- \cos(\theta_{\textbf{i}+\hat{\boldsymbol\mu}_j}-\theta_{\textbf{i}})],\label{eq:F_phi}
\end{equation}
where $d$ is the dimension of a lattice, $\hat{\boldsymbol\mu}_j$ is the $d$-dimensional unit vector in the $j$-th direction,
i.e.~$\hat{\boldsymbol\mu}_1=(1,0,\ldots,0)$, $\hat{\boldsymbol\mu}_2=(0,1,0,\ldots,0)$, etc. Moreover,
$\textbf{i}$ stands for the
lattice coordinate $(i_{1},\dots,i_{d})$. Here, the sum runs over all $i_{1},\dots,i_{d}$ each
running from $1$ to $N$, and each $\theta_{\textbf{i}}\in(-\pi,\pi]$.
$d$ is the dimension of the lattice, and $N$ is the number of sites for each dimension,
so the number of $\theta$ values required to specify the configuration is $N^d$.
Because of the $\theta_{\textbf{i}+\hat{\boldsymbol\mu}_j}$ terms in the model, 
we have to impose a boundary condition, which can be written as
$\theta_{\textbf{i}+N\hat{\boldsymbol\mu}_j}=(-1)^{k}\theta_{\textbf{i}}$ for $1\le j\le d$,
where $N$ is the total number of lattice sites in each dimension. 
In the present work we choose periodic boundary conditions (PBC) specified by $k=0$. 
To remove the global O($2$) degree of freedom due to the 
symmetry $\theta_{\textbf{i}}\to\theta_{\textbf{i}}+\alpha,\forall \textbf{i}$,
where $\alpha$ is an arbitrary constant angle, we fix one of the variables to zero: 
$\theta_{(N,N,\ldots,N)}=0$.

All the SPs of the 1D XY model were found analytically in Ref.~\cite{Mehta:2010pe} for PBC and in 
Refs.~\cite{Mehta:2009,vonSmekal:2007ns} for APBC. 
In 2D, for the $3\times 3$ case, all the isolated SPs were characterised numerically 
using the NPHC method in Refs.~\cite{Mehta:2009zv,Hughes:2012hg}
For larger lattices in 2D, only a few classes of SPs 
were found in Refs.~\cite{2011PhRvL.106e7208C,Hughes:2012hg,Nerattini:2012pi,Mehta:2013iea}
The model is bounded from above and below, and
complete knowledge about the potential energy landscape beyond the $3\times 3$ lattice is yet to be achieved.
Hence, this model is as an ideal testing-ground for comparing the 
methods described in the previous Section, while 
providing important new information about the potential energy landscape.

In Table ~\ref{tab:timing}, we present a comparison of the methods 
described in the previous Section for the XY model in 1D and 2D. 
For the relaxation timings quoted in Table~\ref{tab:timing}, two values of
$\delta$ and $\Delta$ have been used. The Table indicates that
the relaxation approach, even without the inversion step, 
can be enormously more efficient than
the usual random search scheme described above. 
The RFI method, which guarantees a more exhaustive
search for maxima of the original potential, can in principla find all
the SPs of every index. 

To find only transition states (TS) and minima, as opposed to finding all the SPs of 
all the indices, one can start the relaxation from 
a low index SP. Table
\ref{tab:timing_ts} presents a comparison of this kind. We
start relaxation from a SP of index $i$, and successively relax until
a minimum is found. For the present systems we find that starting relaxation from SPs of index
two samples the TS and minima slower than starting relaxation from SPs
of index
three (but still faster and more systematically than the random search
method). This result is a consequence of using initial random
searches to find the SPs that we wish to relax. If a SP has a large basin of
attraction, then the random search method can converge to it many
times. Hence, while relaxation did find more TS and minima in less
time than the random search method, it is possible to significantly improve the
gains by sampling the SPs of index two more
efficiently. Improved sampling of the index two SPs can be achieved 
using multiple values for the step size, $\Delta$, in the random
search routine with at least one relatively
big value, so that it is possible to move out of a large
basin. Alternatively, starting relaxation from SPs of index three
should sufficiently sample the SPs of index two. Both approaches were
found to 
give comparable results. In Table \ref{tab:timing_ts}, we compare
sampling minima and TS using the random
search method against relaxation from SPs of index three. We find that
relaxation is significatly more efficient, both in terms of speed and
improved sample size, when restricting our search to just TS and minima. 

Another advantage of the RFI method, apart from being able to find all the SPs, is
that it is less dependent on the random initial guesses. The RFI approach only relaxes 
energy distinct SPs and as such has a rejection rate before optimisation, in contrast to the random search method. 
Of course, the RFI method still employs random searches to find distinct maxima
in the first step. 
Hence if any maximum is missed it will not necessarily find {\it all} the SPs but will 
probably still perform much better than the random search method in sampling the lower index SPs. 
Using the interval method to find maxima combined with RFI would in principle find all maxima and all lower index saddles. 
Moreover, in practice, it was observed that in certain cases the relaxation
resulted in an incomplete sampling of the lower index saddles (but
still a larger sample than the random search method). This result was probably
due to the lack of numerical stochasticity. Stochasticity can be
introduced using a range of values for the magnitude of the
perturbation away from a SP, $\delta$, combined with different maximum step sizes, $\Delta$,
for the subsequent Newton-Raphson/eigenvector-following geometry optimisation.

\section{Conclusion}
Finding SPs of a potential is an important problem in the physical sciences, 
and there are few methods 
that can find all the SPs of every index.
In the present work we propose a novel approach, Relaxation-followed-by-Inversion (RFI), 
based on relaxing the saddles of index $i$ to find all 
the connected saddles of index $i-1$. 
Hence, starting from $i_{\rm max}$ and going all the way down to $i=0$, we can guarantee to find 
all the SPs of all indices, provided that the procedure is run long enough, so that all the maxima of the potential are obtained, and that we have taken care of the choice of geometry optimization parameters, such as step sizes.

In Table \ref{tab:timing}, 
we compare the efficiency of RFI with the random search hybrid eigenvector-following method 
for the XY model without disorder and conclude that RFI is much faster.
In Table \ref{tab:timing_ts}, we perform relaxation from index three
SPs in order to find those of 
index $1$ and $0$, i.e., transition states and minima. Relaxation
proved to be more
efficient at sampling a greater number of index $1$, $0$ SPs in less time. 

This approach can in principle find all the stationary points of a potential that is bounded from above, provided that all the maxima
can be obtained. Almost all continuous spin model Hamiltonians fall into this category. For potentials that are not bounded 
from above, the relaxation phase starting from index $3$ or $4$ saddles instead of maxima can be employed to find transition states and minima. 
This approach 
is again shown to be far more efficient than searching from randomly chosen configurations.

\section*{Acknowledgements}
DJW and DM gratefully acknowledge support from the EPSRC~and~the~ERC. 
DM was also supported by a DARPA Young Faculty Award.
CH acknowledges support from STFC and the Cambridge Home and European Scholarship scheme. 


\end{document}